\documentclass[a4paper]{article}
\usepackage{Odyssey2020}
\usepackage{epsfig,amssymb,amsmath}
\usepackage{comment}
\usepackage{hyperref}
\usepackage{cite}
\usepackage{url}
\ninept

\usepackage{multicol}
\usepackage{multirow}
\usepackage{tabularx}
\usepackage{booktabs} % To thicken table lines

\title{An empirical analysis of information encoded in disentangled neural speaker representations}
\makeatletter
\def\name#1{\gdef\@name{#1\\}}
\makeatother
\name{\em Raghuveer Peri, Haoqi Li, Krishna Somandepalli, Arindam Jati, Shrikanth Narayanan}
\address{Signal Analysis and Interpretation Laboratory,  \\
University of Southern California, \\
Los Angeles, CA, USA. \\
\{rperi, haoqili, somandep, jati\}@usc.edu, shri@ee.usc.edu }
\begin{document}
\maketitle
\begin{abstract}
%<UNFINISHED>
%Characteristics of robust speaker representations.. (invariant to nuisance factors, reliable for speaker factors)
The primary characteristic of robust speaker representations is that they are invariant to factors of variability not related to speaker identity. 
%Robustness in speaker representations has been extensively studied in literature. 
Disentanglement of speaker representations is one of the techniques used to improve robustness of speaker representations to both intrinsic factors that are acquired during speech production (e.g., emotion, lexical content) and extrinsic factors that are acquired during signal capture (e.g., channel, noise). 
%The goal in obtaining disentangled neural speaker representations is to separate speaker-related factors from nuisance factors that are unrelated to speaker characteristics. 
Disentanglement in neural speaker representations can be achieved either in a supervised fashion with annotations of the nuisance factors (factors not related to speaker identity) or in an unsupervised fashion without labels of the factors to be removed. In either case, it is important to understand the extent to which the various factors of variability are entangled in the representations. In this work, we examine speaker representations with and without unsupervised disentanglement for the amount of information they capture related to a suite of factors. 
%We employ deep neural network based classification models to investigate the amount of information related to various factors. 
Using classification experiments we provide empirical evidence that disentanglement reduces the information with respect to nuisance factors from speaker representations, while retaining speaker information. This is further validated by speaker verification experiments on the VOiCES corpus in several challenging acoustic conditions. We also show improved robustness in speaker verification tasks using data augmentation during training of disentangled speaker embeddings. Finally, based on our findings, we provide insights into the factors that can be effectively separated using the unsupervised disentanglement technique and discuss potential future directions.
\end{abstract}

\section{Introduction}
\label{sec:intro}
%Speaker recognition involves extracting low-dimensional representations of speech that can reliably identify a speaker. There are numerous applications of automatic speaker recognition, 
Speaker embeddings are low-dimensional representations of speech that capture speaker characteristics. They have numerous applications in tasks involving automatic speaker recognition, such as speaker diarization (identifying who spoke when)~\cite{beigi2011speaker}, voice biometrics \cite{10.1145/348941.348995}, anti-spoofing \cite{anti-spoof} and personalized services such as in smart home devices \cite{smart-home}. Real-world speaker recognition requires that speaker embeddings capture speaker characteristics robust to all other attributes unrelated to the speaker's identity, such as the acoustic conditions, microphone characteristics and (aspects of) lexical content in the speech signal.

The topic of extracting robust speaker representations invariant to various factors of variability has been widely studied in the literature. A powerful joint factor analysis~(JFA) based approach was developed in \cite{kenny2007joint}, where speaker `supervectors` were factored into speaker-independent, speaker-dependent, channel-dependent and residual factors. The goal was to separately model the different factors of variability. But it was found that the channel-dependent factors, which were expected to capture only the characteristics of the transmission channel, also contained speaker-related information \cite{10.5555/1751362}. To overcome this challenge, a total variability modeling~(TVM) approach was proposed \cite{5545402}, where no distinction was made between the speaker and session variability factors. These speaker embeddings, called i-vectors are further processed through additional channel compensation steps in the total variability space to minimize session variability \cite{solomonoff2007nuisance, Castaldo2007CompensationON}. They have been shown to perform well on speaker verification tasks.
%was modeled in a single total variability space, in which speech utterances were represented as low-dimensional embeddings called i-vectors. 
%By design, i-vectors capture the modes of highest variability, which can contain speaker-related variability along with all other factors unrelated to the speaker identity. Additional channel compensation steps were developed to minimize the channel information while retaining the speaker information from i-vectors \cite{solomonoff2007nuisance, Castaldo2007CompensationON}.

Recently, supervised speaker modeling techniques have been developed \cite{variani2014deep, snyder2018x}. These methods differ from the previous approaches in that they do not try to explicitly separate the factors of variability. The speaker representations proposed in \cite{snyder2018x}, called x-vectors, are extracted from the bottleneck layer of a time-delay neural network, which was trained on a large corpus of augmented audio recordings to recognize speaker identity. 
%These audio recordings were further augmented by artificially adding background noise and music at varying signal-to-noise ratio levels. In order to simulate the effect of reverberation, the audio signals were convolved with various room impulse responses. 
They have been shown to outperform i-vectors for speaker verification, especially in utterances that are shorter than 10 seconds \cite{snyder2017deep}. Systems employing x-vectors have achieved state-of-the-art performance in applications such as speaker verification \cite{snyder2018x} and speaker diarization \cite{7953094}. Since x-vector systems were not trained to explicitly remove factors of variability, they retain information unrelated to the speaker identity \cite{raj2019probing, Williams2019DisentanglingSF}.
%Numerous studies have tackled the problem of robust speaker recognition using a variety of techniques. In the earlier works, signal domain speech enhancement and microphone array processing techniques for speaker recognition were explored \cite{607754,596134, mccowan2001robust}. But those techniques assume prior knowledge of the noise, and the assumption fails to hold in practical applications including non-stationary noise conditions. Feature domain approaches to induce robustness were investigated in many other works \cite{mammone1996robust,reynolds2003channel, pelecanos2001feature}. 

More recently, other approaches to obtain robust speaker embeddings have been proposed, where models were trained to induce robustness to specific factors of variability in a supervised fashion using additional labels of channel conditions \cite{8682064,zhou2019training}. However, it is not practical to obtain such labelled data in real-world scenarios where the recording conditions or communication context are often unknown. In order to overcome this challenge, we recently proposed a method to disentangle speaker-related factors from the other factors unrelated to speaker identity without prior knowledge of the channel conditions \cite{Peri2019RobustSR}.
%extract robust speaker embeddings 
%without prior knowledge of the factors present in the signal unrelated to speaker identity \cite{Peri2019RobustSR}. 
The speaker embeddings proposed in \cite{Peri2019RobustSR} were obtained from x-vectors using an adversarial invariance induction technique. We showed improvements on speaker verification tasks using this approach of disentangling speaker-related factors from nuisance factors in the embeddings, without explicit supervision of the factors to be disentangled.

However, analysis of the effect of disentanglement on speaker representations has been under-explored. It is important to understand the extent to which the various factors are entangled in these representations to compare the performance of different speech embeddings for downstream tasks.
Such analyses also provide valuable directions to improve the task-specific robustness of speech representations to various confounding factors. In this work, we aim to understand the effect of disentanglement on the amount of information retained in the speaker embeddings with respect to various factors.
\vspace{5pt}\\
The contributions of this work are as follows:
\begin{itemize}
    \item Identify speech datasets to analyze the information encoded in speaker representations with respect to a suite of possible factors including speaker identity, gender, lexical content, emotion, sentiment, noise-type and recording channel.
    %\item Investigate the effect of disentanglement on the amount of information encoded with respect to the various factors in speaker embeddings.
    \item Assess the extent to which the speaker-related factors and factors unrelated to speaker identity are encoded in speaker embeddings with and without disentanglement, where disentanglement is achieved using the method proposed in \cite{Peri2019RobustSR}.
    %\item Assess the effect of disentanglement of the speaker-related factors from the other factors in speaker embeddings to understand the extent to which each factor is encoded in th
    %\item Analyze the effect of artificial data augmentation in models trained to disentangle speaker representations.
    \item Show benefits of augmenting the data used for training the disentanglement models in further improving the robustness of speaker recognition.
\end{itemize}

\begin{table}
\caption{Factors in speaker embeddings and corpora used}
\centering
\scalebox{0.86}{\begin{tabular}{c|c|c}
\toprule
                                 & \begin{tabular}[c]{@{}c@{}}Factors\\ considered\end{tabular} & \begin{tabular}[c]{@{}c@{}}Study corpora\end{tabular}             \\ \midrule
\multirow{3}{*}{Channel factors} & Mic                                                          & \multirow{3}{*}{VOiCES\cite{richey2018voices}}                                             \\
                                 & Noise                                                        &                                                                     \\
                                 & Room                                                         &                                                                     \\ \midrule
\multirow{3}{*}{Content factors} & Emotion/setiment                                             & IEMOCAP\cite{busso2008iemocap}, MOSEI\cite{mosei}                                                      \\
                                 & Language                                                     & Mozilla\cite{ardila2019common}                                                             \\
                                 & Lexical                                                      & RedDots\cite{lee2015reddots}                                                             \\ \midrule
\multirow{2}{*}{Speaker factors} & \begin{tabular}[c]{@{}c@{}}Speaker identity\end{tabular}   & \begin{tabular}[c]{@{}c@{}}IEMOCAP, VOiCES, \\ RedDots\end{tabular} \\
                                 & Gender                                                       & IEMOCAP                                                             \\ \bottomrule
\end{tabular}\label{table:factors}}
\end{table}

\section{BACKGROUND}
\label{sec:background}
%Talk about 3 factors and related work. 
%Table with factors, example corpora, related work
%As we have mentioned in Section \ref{sec:intro}, several works exist in literature that have attempted to understand and tackle the various factors of variability present in speech.
Several works have studied the effect of intrinsic factors which are acquired by the signal during the speech production stage, such as the emotional content, lexical information, language \cite{intra-speaker, 7953216, bao2007emotion}. Other studies have focused on extrinsic factors that are acquired at the signal recording stage \cite{nandwana2018robust}.
%These factors arise either due to the intra-speaker variability in prosody or the spoken content of the particular utterances \cite{intra-speaker, 7953216, bao2007emotion}. 
These different factors of variability are entangled to varying degrees in the speaker representations.

Following \cite{wang2017does}, for ease of analysis, we categorize the factors into three distinct categories: \textit{channel factors} which are encoded in the speech signal during the process of signal acquisition, such as background acoustic noise, microphone characteristics, room response and acoustic scene, \textit{content factors} which encode information about the spoken content including its prosodic content, such as the lexical aspects, emotion, sentiment, language and accent \cite{8462169}, \textit{speaker factors} which are inherent to the speaker, such as speaker identity, age and gender.

In Table~\ref{table:factors} we summarize few of the exemplar factors belonging to these three categories. We also list the corpora that we have considered in this paper. In the following sections, we discuss the $3$ different factors of variability in detail.
%As shown in Table \ref{table:factors}, several previous works have attempted to understand the information encoded in speaker embeddings.
\vspace{5pt}\\
\textbf{Channel factors}: These factors have been extensively studied in literature in the context of speaker recognition. The goal of robust speaker recognition systems is to rid the speaker embeddings of these factors, while retaining the speaker-related factors. As mentioned in Section \ref{sec:intro}, early speaker recognition systems either attempted to decompose channel-dependent factors from the speaker-dependent factors during speaker modeling \cite{kenny2007joint}, or introduced channel compensation steps in the total variability space \cite{Castaldo2007CompensationON}.
%While early JFA-based techniques attempted to decompose channel-dependent factors from the speaker-dependent factors, i-vector based methods capture all the factors of variability in a single low-dimensional total variability space and the separation of channel factors was done through additional channel compensation methods . 
X-vector based methods do not make this separation explicit and can contain significant amount of channel-related information, as shown through channel classification experiments \cite{raj2019probing}.
%The goal of robust speaker recognition systems is to strip the speaker embeddings of these factors, while retaining information of speaker factors.
\vspace{5pt}\\
\textbf{Content factors}:
They encode information about the spoken content including the prosodic variations in speech. Examples include lexical content, emotional content, sentiment, language identity etc.
%A person's emotion state and speaSRE UAI Odyssey V3king style along with lexical content of an utterance often manifest as variabilities in speech. 
These factors have been shown in past works to be entangled with speaker-related factors in speaker representations. For example, examining the amount of phoneme-level information present in speaker embeddings, it was found that better performance in phoneme classification tasks does not translate to improved speaker recognition performance \cite{lozano2016analysis}. Furthermore, speaker embeddings that perform well for speaker recognition tasks have been shown to capture segment-level characteristics rather than low-level phoneme information \cite{Shon2018FrameLevelSE}.

%The various factors encoded in i-vectors and x-vectors, have been previously explored \cite{wang2017does,raj2019probing}. 
It was found in past literature that both i-vectors and x-vectors capture speaker-related factors along with significant information related to the other factors \cite{wang2017does,raj2019probing}. However, it is important to note that the inter-dependence between factors arising due to the construction of the dataset were not considered in \cite{raj2019probing}.
%But we found that some of the experiments in \cite{raj2019probing} had not considered the inter-dependence between variables. 
For example, lexical content information was investigated using sentence classification task, but the data included sentences that were unique to particular speakers, making it challenging to disambiguate the speaker information from the lexical information using only the sentence classification performance.
Williams et al.~\cite{Williams2019DisentanglingSF}, examined x-vectors for the task of subject-independent emotion recognition. Their experiments showed that x-vectors trained for speaker recognition task could predict the emotion content in speech even without any additional supervision using emotion labels.
%the emotional content encoded in x-vectors was analyzed, but the goal there was to retain emotion information, while discarding speaker information. 
%Language identity is another important aspect of speech, and speaker representations are expected to capture speaker characteristics irrespective of the language spoken. 
%There have been several works that perform language identification using TVM-based approach \cite{dehak2011language,lopez2014automatic}. Although TDNN architecture similar to that of x-vectors has been developed for language id \cite{snyder2018spoken}, in the context of speaker recognition tasks, to the best of our knowledge, the language-related information encoded in the supervised speaker embeddings has not been explored well so far.
%Language classification performance can be treated as a proxy of the amount of information pertaining to the spoken language
\vspace{5pt}\\
\textbf{Speaker factors}:
\begin{comment}
% These are inherent to the speaker, and include the identity of the speaker, their gender and age. Speaker embeddings should reliably capture these factors. To study the amount of information pertaining to these factors in speaker representations, it is necessary to choose datasets that control for each individual factor of variability separately, which is a challenge. 
% For example, in a dataset consisting of recordings where each speaker speak into their individual microphones, it is possible that a model trained to classify microphone types can provide a very high speaker classification accuracy. This can be handled by ensuring that all other factors of variability are kept constant when studying the speaker factors. (how to rewrite?)
\end{comment}
These factors are inherent to the speaker (c.f., demographics), and capture the identity of the speaker to various degrees. For example, while gender and age are not sufficient to fully recognize a speaker’s identity from speech, robust speaker embeddings can effectively capture these dimensions. Thus examining how speaker embeddings perform for identifying these individual factors is key to understanding robustness for speaker recognition. In particular, it has been found that i-vectors and x-vectors can be successfully used to classify gender \cite{wang2017does, raj2019probing}. However, it is still unclear whether speaker embeddings that are more discriminative of gender significantly improve speaker recognition performance. In this paper we also assess these factors when the labels are available.
\vspace{5pt}\\
We probe into the information encoded in two speaker embeddings based on the work in \cite{Peri2019RobustSR} and compare with that of x-vectors. We show through speaker verification experiments on the VOiCES dataset \cite{richey2018voices} that the speaker embeddings after disentanglement retain information pertaining to speaker identities. Since the speaker representations extracted using the technique in \cite{Peri2019RobustSR} do not rely on labelled information about the nuisance factors, we hypothesize that such disentanglement results in speaker representations that contain minimal information related to all the nuisance factors, while retaining speaker-related information.
%, while minimizing information related to all other factors of variability. 
We chose classification tasks related to a particular factor of variability as a proxy to the amount of information encoded in the speaker representations related to that factor as has been done in several previous works \cite{wang2017does, Shon2018FrameLevelSE, adi2016finegrained, Williams2019DisentanglingSF}. Results suggest that the process of disentanglement reduces the amount of information present in speaker embeddings pertaining to factors not related to the speaker identity.
%\cite{wang2017does, raj2019probing, Williams2019DisentanglingSF, lozano2016analysis, Shon2018FrameLevelSE}.

% \begin{figure}[!t]
%   \centering
%   \centerline{\includegraphics[width=6cm]{drawing.png}}
% %  \vspace{2.0cm}
%   %\centerline{(a) Result 1}\medskip
% \caption{Unsupervised adversarial invariance applied for speaker recognition}
% %\caption{Adversarial invariance architecture}
% \label{fig:uai}
% %
% \end{figure}

\section{METHODOLOGY}
\label{sec:methods}
%\subsection{Feature extraction}
%\label{ssec:x-vector}
%The goal of our work is to induce invariance in speaker embeddings to various sources of variability. 
%The goal of this work is to investigate the amount of information related to the different factors encoded in speaker embeddings with and without disentanglement. 

\subsection{Baseline}
\label{ssec:x-vector}
We extract x-vectors using the publicly available pre-trained model~\footnote{https://kaldi-asr.org/models/m7} and consider them as the baseline speaker embeddings trained without disentanglement. The x-vector model was trained on a large corpus of audio recordings to predict speakers. These audio recordings were further augmented by artificially adding background noise and music at varying signal-to-noise ratio levels \cite{snyder2018x}. In order to simulate the effect of reverberation, the audio signals were convolved with various room impulse responses. As discussed in Section \ref{sec:background}, since x-vectors were not trained with an explicit disentanglement stage, they retain information pertaining to factors unrelated to speaker identity.

\subsection{Disentangled speaker embeddings}
\label{ssec:UAI}
We employ the unsupervised adversarial invariance~(UAI) technique, which was originally proposed in \cite{jaiswal2018unsupervised} and modified to disentangle the speaker factors from the other factors present in x-vectors \cite{Peri2019RobustSR}. The central idea behind the UAI technique is to project the input x-vectors into a split representation consisting of two embeddings, referred to as $\mathbf{h}_{1}$ and $\mathbf{h}_{2}$. While $\mathbf{h}_{1}$ is trained with the objective of capturing speaker-specific information, $\mathbf{h}_{2}$ is trained to capture all other nuisance factors. This is done by training two models in an adversarial fashion. The goal of one model, called main model, is to predict speakers using $\mathbf{h}_{1}$ as input and reconstruct the x-vectors using $\mathbf{h}_{2}$ as input. The second model, called the adversarial model, is trained to minimize the predictive power of $\mathbf{h}_{1}$ from $\mathbf{h}_{2}$ and vice-versa. The two models are adversarially trained in a minimax game fashion. The speaker prediction task forces $\mathbf{h}_{1}$ to capture speaker-related information, while the reconstruction task ensures that $\mathbf{h}_{2}$ captures information related to all factors.
Further, a noisy version of $\mathbf{h}_{1}$ is used during the reconstruction task along with $\mathbf{h}_{2}$, so that the network learns to treat $\mathbf{h}_{1}$ as an unreliable source of information for the reconstruction, hence ensuring $\mathbf{h}_{1}$ does not contain information about factors other than the speaker. Detailed explanation of the UAI method to disentangle speaker representations can be found in \cite{Peri2019RobustSR}.
%have been shown to capture information of factors not related to speaker identity.
%Our objective, in using unsupervised adversarial invariance ~(UAI) technique, is to disentangle the speaker factors from the other factors present in x-vectors.
%Hence we set up several experiments to investigate the amount of information related to the different factors encoded in the embeddings extracted using our system and compare with that in x-vectors.
%We hypothesize that since our models were trained to retain speaker information while discarding the information irrelevant to the speaker prediction task, our embeddings should capture less amount of information regarding the channel and content factors.

We trained two models using the UAI technique that differ in the training data used:
\begin{itemize}
    \item M1: Trained without artificially augmented data
    \item M2: Trained with artificially augmented data
%\item M3: Trained with Vox-aug-sub
\end{itemize}
%The model M1 is similar to the one proposed in \cite{Peri2019RobustSR}. We developed the model M2 to investigate the effect of introducing additional variability in the training samples by adding artificial augmentations on the information encoded in speaker embeddings.
For consistency, we use the same model as proposed in \cite{Peri2019RobustSR}, denoted M1. We developed the model M2 to investigate the benefit of artificially augmenting the training samples on the robustness of speaker embeddings.
%These audio recordings were further augmented by artificially adding background noise and music at varying signal-to-noise ratio levels. In order to simulate the effect of reverberation, the audio signals were convolved with various room impulse responses.

\subsection{Factor prediction}
\label{ssec:Probe}
As mentioned in Section \ref{sec:background}, to measure the extent of entanglement of the various factors in speaker embeddings, we designed classification experiments. In these experiments, the speaker embeddings were used as input to predict each factor of variability. Such a method of analysis assumes that if a factor is encoded in the speaker embeddings, a classifier can be trained to predict the factor using the speaker representations as input. Further, the classification accuracy can be considered a proxy for how well the factor has been encoded in these speaker embeddings \cite{wang2017does}.

\begin{figure}[t]

  \centering
  \centerline{\includegraphics[width=7cm,height=5cm]{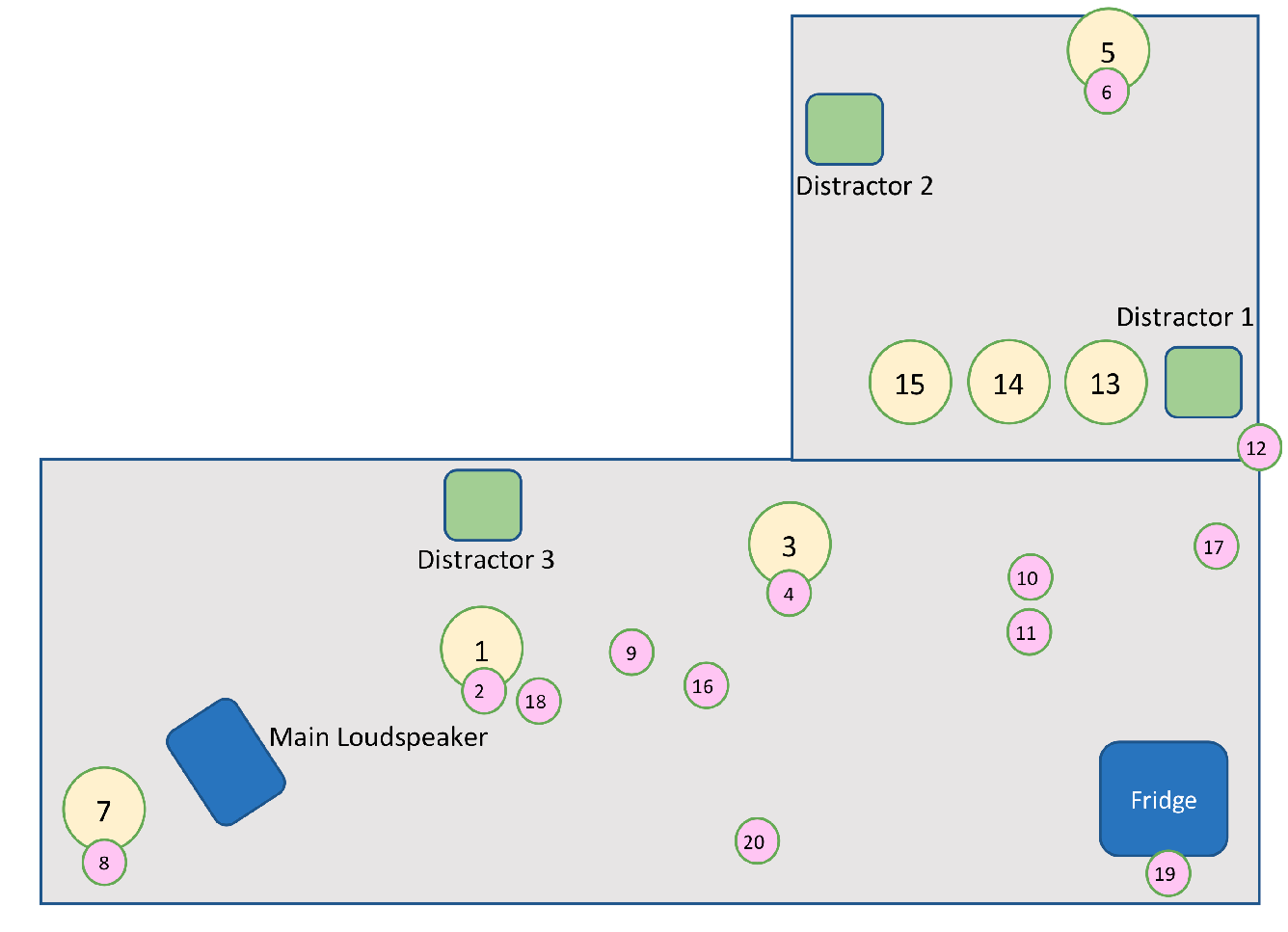}}
\caption{Example room configuration in VOiCES dataset\cite{nandwana2019voices}. Distractor represents noise source and circles represent microphones}
\label{fig:voices}
\end{figure}
\vspace{-0.05in}

\begin{figure*}[!t]
  \centering
  \centerline{\includegraphics[width=16cm,height=7cm]{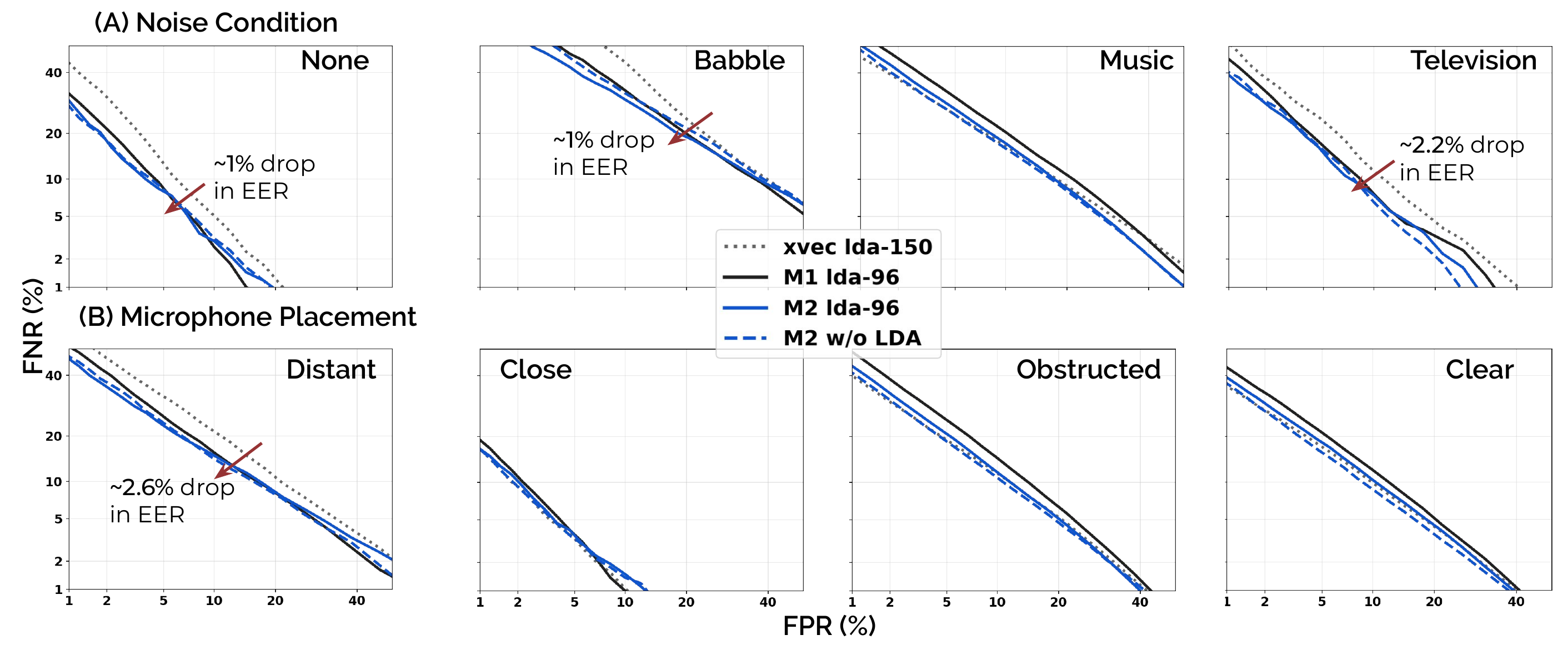}}
  %[width=7cm,height=5cm]
\caption{DET curves of speaker verification task using different speaker embeddings with and without disentanglement in several (A) Noise conditions and (B) Microphone placements.}
\label{fig:det}
\end{figure*}

\section{DATASETS}
\label{sec:datasets}
As mentioned in Section \ref{sec:background}, we performed experiments on a number of publicly available datasets. Each dataset was chosen to enable the exploration of individual factors, while providing an analysis of how these factors are manifested in speaker representations on real-world data.
\vspace{5pt} \\
\textbf{VOiCES:}
%\subsection{VOiCES}
Recordings collected from $4$ different rooms with microphones placed at various fixed locations, while a loudspeaker played clean speech samples from the Librispeech \cite{7178964} dataset. Along with speech, noise was played back from loudspeakers present in the room, to simulate real-life recording conditions. Figure~\ref{fig:voices} shows one such room configuration and data collection setup where "Distractor" represents noise source and the circles represent the available microphones. This dataset was chosen for the availability of  annotations regarding the acoustic conditions including noise types and microphone locations. We use the VOiCES corpus to evaluate speaker verification performance under various acoustic conditions. We perform our evaluations on the phase-2 release of the dataset to be consistent with the data used during the evaluation in VOiCES challenge \cite{nandwana2019voices}. This subset consists of recordings from $2$ rooms collected using $20$ microphones and contains $4$ distinct noise types.
We also use this dataset to explore the amount of information related to the speaker, room, noise and microphone type.
%We divide our analysis into two subsets: one that was provided as the evaluation set in the VOiCES challenge \cite{nandwana2019voices} called \textbf{voices-eval} and the second consisting of recordings released during phase 2, called \textbf{voices-phase2}. This was done 
\vspace{5pt} \\
\textbf{IEMOCAP:}
%\subsection{IEMOCAP}
A multi-modal corpus consisting of video, audio, face motion, hand movements and text transcriptions. These modalities were recorded from $10$ actors during scripted and enacted conversations. Annotations are provided for discrete and continuous emotion ratings. Following \cite{Williams2019DisentanglingSF}, we make use of a small subset consisting of the audio portion of the data with a single emotion label per utterance, leading to $1403$ utterances corresponding to $4$ discrete emotions, angry, sad, happy and neutral. This dataset provides the capability to investigate the amount of emotional information present in the speaker embeddings.
Further, we use this dataset to perform speaker identification experiments to explore the speaker information present in the embeddings.
%Further, since the number of speakers in the dataset was limited, we performed speaker identification experiments to explore the speaker information present in the speaker embeddings.
%Further, this dataset allowed us to analyze speaker verification performance under various emotions.
\vspace{5pt} \\
\textbf{MOSEI:}
%\subsection{MOSEI}
An audio dataset of more than $23000$ YouTube videos annotated for emotion and sentiment. Audio from this corpus was used for the analysis of emotion and sentiment information captured in speaker embeddings. Since, the emotions elicited in all the recordings were spontaneous, this database was useful to validate the generalizability of our analysis to real-world spontaneous data. We performed pre-processing of the raw labels provided in the corpus. Each audio recording in the dataset was annotated for scores corresponding to six emotion labels. We chose the emotion label with the maximum score for each audio recording. For sentiment analysis, we convert the labels provided in the dataset into $3$ distinct labels, positive, negative and neutral sentiment.
%haoqi: citation for using the dataset this way?
\vspace{5pt} \\
\textbf{Mozilla:}
%\subsection{Mozilla}
A publicly available corpus of crowdsourced audio recordings. It consists of transcribed text from multiple languages, from which we chose a subset of $4$ languages~(English, German, Mandarin and Turkish) for ease of analysis. The languages are chosen such that sufficient number of samples exist for each language. This corpus is used for the language prediction task.
\vspace{5pt} \\
\textbf{RedDots:}
%\subsection{RedDots}
Speech recordings collected from participants using their own audio recording device, typically a mobile phone. The dataset comprises recordings of participants reading out sentences, some of which were common across all the participants while the others were unique to each participant. To ensure that experiments involving lexical content factor were not confounded by variability due to speakers, we prune the data and obtain a small subset of recordings consisting of only the $10$ common sentences spoken by all the speakers. This corpus is used for analysis of lexical content and also for the gender prediction task.
%To ensure that analysis performed to compute the amount of lexical content information present in the speaker embeddings is not corrupted by variabilities due to speakers, we prune the data and obtain a small subset of recordings consisting of only the $10$ common sentences spoken by all the speakers.
%Annotations of gender are available, which were used to analyze gender-related information encoded in speaker embeddings. 
%We further prune the data and obtain a small subset of recordings consisting of only the $10$ common sentences spoken by all the speakers.
%This way we ensure that analysis performed to compute the amount of lexical content information present in the speaker embeddings is not corrupted by variabilities due to speakers
\vspace{5pt} \\
\textbf{Vox\textunderscore clean:}
%\subsection{Vox\textunderscore clean}
To ensure fair comparison with x-vectors baseline, we trained the models in \cite{Peri2019RobustSR} using a subset of the data that was used to develop the pre-trained x-vector models. It consists of a combination of the development and test splits of VoxCeleb2 \cite{chung2018voxceleb2} and the development split of VoxCeleb1 \cite{nagrani2017voxceleb} datasets. 
%It consists of \textit{in-the-wild} recordings annotated for speakers. 
Since the dataset is sourced from unconstrained recording conditions, there exists a huge amount of variability in terms of channel conditions and the content factors such as lexical and emotion content \cite{albanie2018emotion}. It consists of audio recordings corresponding to $1.2M$ utterances from $7323$ distinct speakers. We refer to this subset as Vox\textunderscore clean.
% %Start same as ICASSP
% The training data for our models in \cite{Peri2019RobustSR} consists of a combination of the development and test splits of VoxCeleb2 \cite{chung2018voxceleb2} and the development split of VoxCeleb1 \cite{nagrani2017voxceleb} datasets. This is consistent with the split that was used to train the pre-trained Introduction, Background, Datasetsx-vector model (mentioned in Section \ref{ssec:x-vector}). It consists of speaker annotated \textit{in-the-wild} recordings from celebrity speakers. As such the dataset is sourced from unconstrained recording conditions. For brevity, henceforth, we refer to this subset of the VoxCeleb dataset as Vox. It consists of audio recordings corresponding to $1.2M$ utterances from $7323$ distinct speakers.
% % End same as ICASSP
We also augment this data by artificially adding noise following \cite{snyder2018x}, using music and noise samples from the MUSAN corpus \cite{Snyder2015MUSANAM}. Artificial reverberation was simulated using room impulse response samples from \url{https://www.openslr.org/}. We refer to this augmented dataset as \textbf{Vox\textunderscore aug}. This process doubled the amount of training data to $2.4M$ utterances keeping the number of speakers the same.

\section{EXPERIMENTS}
\label{sec:experiments}
\begin{table*}[ht!]
\centering
\caption{Accuracy (\%) of speaker embeddings to classify different factors. Robust speaker embeddings are expected to perform poorly for classifying non-speaker related tasks. The best model for each factor with respect to this criterion is shown in bold}
\centering\vspace{0.1in}
\begin{tabular}{l|l|l|c|c|c}
\toprule
                                 & Factors                  & Dataset                  & x-vector (512 dim) & x-vector + PCA (128 dim) & Ours: M1 (128 dim) \\
                                 \midrule
\multirow{3}{*}{Channel factors} & Room                     & \multirow{3}{*}{VOiCES}           & 99.8               & 99.7                     & \textbf{97.3}               \\
                                 & Mic                      &                                    & 91.0                 & 83.3                     & \textbf{57.4 }              \\
                                 & Noise                    &                                    & 94.9               & 92.2                     & \textbf{76.2}               \\
                                 \midrule
\multirow{5}{*}{Content factors} & \multirow{2}{*}{Emotion} & IEMOCAP                           & 91.5               & 91.7                     & \textbf{80.9}               \\
                                 &                          & MOSEI                             & 62.4               & 61.5                     &\textbf{ 60.4}               \\
                                 & Sentiment                & MOSEI                             & 53.9               & 53.0                       & \textbf{49.3}               
                                                \\
                                 & Lexical                  & RedDots                           & 98.0                 & 97.1                     & \textbf{80.3}               \\
                                 & Language                 & Mozilla                          & 97.2               & 96.6                     & \textbf{95.3}
                                 \\
                                 \midrule
\multirow{4}{*}{Speaker factors} & \multirow{2}{*}{Speaker} & VOiCES                            & \textbf{99.6}               & 99.6                     & 98.5               \\
                                 &                          & IEMOCAP                           & 80.1               & \textbf{81.8}                     & 77.8               \\
                                 & \multirow{2}{*}{Gender}                   & IEMOCAP                           & \textbf{98.9}               & 98.0                     & 97.2               \\
                                 &                   & RedDots                           & 99.0                 & \textbf{99.5}                     & 97.6 \\
                                 \bottomrule
\end{tabular}\label{table:probing}
\end{table*}
\subsection{Speaker verification}
\label{ssec:verification}
\subsubsection{Setup}
%We performed several experiments on the VOiCES dataset to verify the efficacy of our method in extracting speaker embeddings that are invariant to various factors of variability.
We performed several experiments on the VOiCES dataset to verify the efficacy of disentanglement in making speaker embeddings robust to various factors of variability. Following \cite{richey2018voices} and \cite{nandwana2018robust}, we experimented with several combinations of noise types and microphone locations to understand the effect of each factor on speaker verification performance. In a similar fashion as \cite{Peri2019RobustSR}, we consider two distinct factors, noise conditions: none, babble, television and music, and microphone location: distant, close, clear and obstructed. By abuse of notation, we denote the disentangled embeddings by the same name as the model from which they were extracted, i.e., M1 and M2.

%We further distinguish between the recordings collected at $2$ different microphone locations (far-mic vs. near-mic) while examining the performance in noisy conditions.
% start same as ICASSP
We apply dimensionality reduction using linear discriminant analysis~(LDA) and score the speaker verification trials using a probabilistic linear discriminant analysis~(PLDA) backend for all the embeddings obtained with and without disentanglement. The LDA and PLDA models were learnt on the training data for our proposed system, while for the baseline x-vector system we used the pre-trained models. For the embeddings extracted after disentanglement (both M1 and M2), we use a dimension of $96$ after LDA (chosen based on speaker verification performance on the VOiCES challenge development portion), while for x-vectors we use $150$ as the reduced dimension to be compatible with the pre-trained PLDA model. We also experimented with removing the LDA-based dimensionality reduction stage on the disentangled speaker embeddings.
% end  same as ICASSP
\subsubsection{Results}
Figure~\ref{fig:det} shows the detection error trade-off~(DET) plots of the various speaker verification experiments with False Positive Rate~(FPR) on the x-axis and False Negative Rate~(FNR) on the y-axis.
Each subplot compares four speaker embeddings,
% \begin{itemize}
%     \item \textbf{xvec lda-150}: x-vector with dimension reduced to $150$ using LDA
%     \item \textbf{M1 lda-96}: speaker embedding from model M1 reduced to $96$ dimensions
% \end{itemize}
x-vector reduced to $150$ dimensions using LDA~(\textbf{xvec lda-150}), M1 reduced to $96$ dimensions~(\textbf{M1 lda-96}), M2 reduced to $96$ dimensions~(\textbf{M2 lda-96}) and M2 without LDA-based dimensionality reduction~(\textbf{M2 w/o lda}).

% Please add the following required packages to your document preamble:
% \usepackage{multirow}

%
The first row of subplots in Figure~\ref{fig:det} shows the DET curves across different noise conditions. We observe that in majority of the cases disentangled speaker embeddings obtain lower error rates at most operating points. In particular, for the television and babble noise conditions, which are considered challenging due their speech-like characteristics \cite{nandwana2018robust}, we observed a respective absolute reduction of 2.2\% and 1\% in the equal error rate~(EER). Furthermore, we observe that for the distant microphone scenario, speaker embeddings after disentanglement provide lower error rates at almost all operating points, with a 2.6\% reduction in EER.
\subsection{Probing analysis}
\label{ssec:probe}
\vspace{5pt}
\subsubsection{Setup}
As mentioned in Section \ref{sec:background}, we analyze the extent of information present in the speaker embeddings with respect to various factors.
%We built $3$ simple feed-forward deep neural network based classifiers, one for each factor.
For each factor that was analyzed, we extracted both x-vectors and disentangled speaker embeddings~(denoted M1) from the corresponding dataset for that factor. We also reduce the dimension of x-vectors~(denoted xvector+PCA) using principal component analysis~(PCA), similar to \cite{Williams2019DisentanglingSF}. This was done to match the dimension of x-vectors with that of the disentangled speaker embeddings, to ensure fair comparison with respect to embedding dimension.
%disambiguate the effect of the embedding dimension on the information encoded.

We trained a classification model for each factor using a simple feed-forward deep neural network. The classifier contains of 4 fully connected layers with 256 hidden neurons and ReLU activation function in between. 
Similar to \cite{Williams2019DisentanglingSF}, we use L2 regularization, Adam optimizer with learning rate lr = 0.0002, and an early-stopping criterion monitored by the loss on validation set.
The classifier models were trained to predict the factor with the speaker embeddings as input.

For the IEMOCAP dataset, for fair comparison, we use the same train and test split as in \cite{Williams2019DisentanglingSF}. For the RedDots dataset, in the lexical content analysis task,  for each speaker we randomly split 80\% of the data into training, 10\% into validation and 10\% as test part. In the gender classification task, based on the previous split, we further perform minority class sub-sampling to balance the data for each gender.

Results summarizing the classification performance of the speaker embeddings with and without disentanglement are reported in Table~\ref{table:probing}, where the accuracy~(\%) values were rounded to the nearest decimal.
\subsubsection{Channel factors}
\label{sssec:channel}
We explore three different channel variability conditions: microphone id, noise and room type. We trained classifiers to predict the microphone at which the recording was collected given the speaker embeddings extracted from a recording. We built classifiers to predict the noise type from the speaker embeddings. We also trained classification models to predict the room in which the recordings were made, using speaker embeddings as input.
\vspace{5pt}\\
\textbf{Results}
\vspace{5pt}\\
 We observe from Table~\ref{table:probing} that there is a fairly large difference in the predictive power of x-vectors and disentangled speaker embeddings (M1) when classifying microphones and noise types. This suggests that the disentanglement method is able to successfully reduce the channel information from speaker representations. However, all embeddings perform well in predicting the room. It is unclear if this behavior is due to an unknown factor of variability between the rooms.

\subsubsection{Content factors}
\label{sssec:content}
Experiments were performed to analyze the information related to four different content factors, lexical content, emotional content, sentiment and language identity using the corresponding corpora mentioned in Table~\ref{table:factors}.
\vspace{5pt}\\
\textbf{Results}
\vspace{5pt}\\
Results on lexical content prediction show an 18\% reduction in classification accuracy upon disentanglement of x-vectors, suggesting the successful reduction in lexical information from speaker embeddings. Results for emotion classification show a 10\% reduction in accuracy on the IEMOCAP dataset. The results on MOSEI dataset show a similar trend of reduced accuracy in M1 for the emotion classification tasks, though the difference is not as substantial. For the sentiment classification task, there is an absolute 5\% reduction in accuracy after disentanglement. Finally, with respect to language prediction, all the embeddings perform well, with a minor decrease in accuracy when disentangled representations are used.

\subsubsection{Speaker factors}
\label{sssec:speaker}
To analyze the speaker-related information present in the speaker representations, we performed separate experiments to predict the speaker identity and speaker gender. 
We used the speaker labels in the VOiCES dataset from which $10$ speakers were randomly chosen for the speaker identification task. We trained classifiers to predict the speaker identity given the speaker embeddings. We also performed similar classification experiments on the IEMOCAP dataset.
%This can be considered as a closed-set speaker identification task, where the test set consists of only the speakers seen during training.
%We used the speaker labels in the IEMOCAP dataset to train classifiers to predict the speaker identity given the speaker embeddings. The dataset consisted of recordings from $10$ speakers. This can be considered as a closed-set speaker identification task, where the test set consists of only the speakers seen during training. 
%We also perform similar classification experiments on the VOiCES dataset from which only $10$ speakers were randomly chosen for the task.
We also trained classifiers on the RedDots dataset to make binary gender predictions using the speaker embeddings as input . %For this purpose, audio data and gender annotations from the RedDots dataset were used.
\vspace{5pt}\\
\textbf{Results}
\vspace{5pt}\\
For the speaker id task on the VOiCES dataset, we observe that M1 performs almost on par with x-vectors, and close to 100\% accuracy, suggesting that speaker identity is retained during the disentanglement process. This result complements the speaker verification performance, see Section \ref{ssec:verification}.

We further observe a similar trend for the gender prediction task on the RedDots dataset, where all the speaker embeddings achieve close to 100\% accuracy. This is consistent with past work \cite{wang2017does, raj2019probing}.

However, results on IEMOCAP did not follow the same trend, though the differences in performance are small. The reason for this could be the implicit co-dependence between the emotions and the speaker labels in the dataset. In order to test this, we conducted hypothesis testing on the contingency table using a chi-squared test\footnote{$H_0$: the two variables of the contingency table are independent. Reject $H_0$ if $p<\alpha$ with $\alpha=0.01$} for the sample frequencies of the emotion and speaker variables in the IEMOCAP dataset. This test~($p\ll 0.001$) revealed that the two variables were dependent in this dataset. This makes it difficult to disambiguate the effects of one factor on the other. Furthermore, we noticed that a few audio recordings 
contained overlapping speech leading to noisy speaker labels.
%contained speech from more than one speaker, leading to noisy speaker labels. This could be another factor that leads to these results.
%This underscores the need to study the relation between the variables/factors that are analyzed in a sample/dataset.
%i.e, amxn table with m speaker and n emotions. This testing revealed that the two variables were depe
%Using Chi-squared test of independence between the emotions and speaker variables, we could reject the null hypothesis that the variables were independent (at a significance level of 0.01), suggesting that the two variables are likely dependent in the subset of data that we considered. 
%This points to the need for joint modeling of factors in a multitask framework.

\subsection{Discussion}
\label{ssec:discuss}
%Based on the experimental results, we provide the following insights into the effect of disentanglement on the factors of variability:
We offer the following insights based on our evaluations using x-vectors and disentangled representations:
\begin{itemize}
    \item Channel factors can be effectively removed from speaker embeddings using unsupervised disentanglement. It further improves robustness of the embeddings for speaker recognition tasks in challenging acoustic conditions. This is consistent with our previous findings 
    \item The emotional content information is also minimized from speaker embeddings during disentanglement, even though emotion labels were not used during training. This can be explained based on the findings in \cite{albanie2018emotion}, where it was shown that audio recordings in the Voxceleb dataset contain expressive speech. 
    %This underscores the need to 
    %We reason that during disentanglement, the emotion content present in training utterances was treated as nuisance factors for the speaker prediction task.
    %Emotions are present in voxceleb \cite{albanie2018emotion}. Since our method explicitly removes all irrelevant information for the speaker recognition task, we can assume that even the emotion information that is present in training utterances is also removed, and this is validated by results on emotion classification.
    \item PCA on x-vectors to reduce dimensionality does not have a significant effect on the prediction performance. This suggests that explicit disentanglement is beneficial over simple dimensionality reduction techniques.
    \item Disentanglement of speaker embeddings retains gender-related information. This suggests that speaker gender, as captured in these representations, is a crucial part of the speaker's identity. If removal of gender factor from speaker representations is desired, then other joint modeling approaches need to be explored. Further, other individual attributes such as age should be investigated in a similar manner.
    \item Consistent with past literature in this domain, we observe that additional data augmentation further improved the performance on the speaker verification task. However, in the context of disentanglement, data augmentation did not reduce the information of the nuisance factors. In our future work, we wish to investigate if informed data augmentation for individual nuisance factor can help disentanglement.
    %Additional data augmentation during disentanglement further improved the performance on the speaker veirification task, consistent with past literature on this domain, about the effect on data augm. However, in the context of disentaglement additional data augmentation did not reduce the information of the nuisancc factors.
    
    %Not reported in this paper, but we found that data augmentation model does not affect factor classification performance. This suggests that although robustness in speaker verification tasks is achieved, in terms of amount of nuisance factor information, data augmentation without supervision of the factors to be removed doesn't help in disentangling the speaker embeddings. Future work can investigate if informed data augmentation for each individual factor would help during disentanglement.

\end{itemize}

\section{Conclusion}
\label{sec:conclusion}
In this work, we showed that disentanglement of neural speaker representations helps improve robustness of speaker embeddings on speaker verification tasks in challenging acoustic conditions. We further showed through experimental evaluation on several datasets and a suite of factors that disentanglement reduces the amount of information encoded in the speaker embeddings that is not related to speaker identity, while retaining speaker-related information.
We hope that these findings will offer novel research directions to develop robust speaker recognition systems.

\bibliographystyle{IEEEbib}
\bibliography{Odyssey2020_BibEntries}

% This could be also done as follows:
%
%\begin{thebibliography}{10}
%\bibitem[1]{aluisio2001learn}Sandra M. Alu\'{i}sio, Iris Barcelos, Jandir Sampaio, and Osvaldo
%N. Oliveira Jr, ``How to learn the many unwritten
%``rules of the game'' of the academic discourse: a hybrid
%approach based on critiques and cases to support scientific
%writing,'' in Proceedings of the IEEE International Conference
%on Advanced Learning Technologies, Madison, USA,
%August 2001, pp. 257–260.
%\bibitem[2]{swales1987writing} John Swales and Hazem Najjar, ``The writing of research
%article introductions,'' Written communication, vol. 4, no.
%2, pp. 175–191, 1987.
%\bibitem[3]{day2012write} Robert Day and Barbara Pastel, How to write and publish
%a scientific paper, Cambridge University Press, 2012.
%\bibitem[4]{teufel2000} Simone Teufel, Argumentative zoning: information extraction
%from scientific text, Ph.D. thesis, University of Edinburgh,
%2000.
%\bibitem[5]{berkenkotter1989social} Carol Berkenkotter, Thomas N. Huckin, and John Ackerman,
%``Social context and socially constructed texts: The
%initiation of a graduate student into a writing research community.
%technical report no. 33.,'' Tech. Rep., Center for
%the Study of Writing, University of California Berkeley \&
%Carnegie Mellon University, 1989.
%\end{thebibliography}

\end{document}